\documentclass{article}
\usepackage{spconf,amsmath,graphicx}

\usepackage{enumitem}
\setlist{nosep, leftmargin=14pt}

\usepackage{float} 

\usepackage{amsmath,amssymb,bm}
\usepackage{booktabs}
\usepackage{graphicx}
\usepackage{float}
\graphicspath{{figs/}}

\usepackage{mwe} 

\title{TopoGate: Quality-Aware Topology-Stabilized Gated Fusion for Longitudinal Low-Dose CT New-Lesion Prediction}
\name{Seungik Cho}
\address{Department of Physics \& Astronomy, Rice University, USA}
\begin{document}
%
\maketitle
\begin{abstract}
	Longitudinal low-dose CT follow-ups vary in noise, reconstruction kernels, and registration quality. These differences destabilize subtraction images and can trigger false “new-lesion” alarms. We present TopoGate, a lightweight model that combines the follow-up appearance view with the subtraction view and controls their influence through a learned, quality-aware gate. The gate is driven by three case-specific signals: CT appearance quality, registration consistency, and stability of anatomical topology measured with topological metrics. On the NLST–New-Lesion–LongCT cohort comprising 152 pairs from 122 patients, TopoGate improves discrimination and calibration over single-view baselines, achieving an area under the ROC curve of 0.65 with a standard deviation of 0.05 and a Brier score of 0.14. Removing corrupted or low-quality pairs, identified by the quality scores, further increases the area under the ROC curve from 0.62 to 0.68 and reduces the Brier score from 0.14 to 0.12. The gate responds predictably to degradation, placing more weight on appearance when noise grows, which mirrors radiologist practice. The approach is simple, interpretable, and practical for reliable longitudinal LDCT triage.
\end{abstract}


\section{Introduction}
\label{sec:intro}
Low-dose CT (LDCT) screening and longitudinal follow-up are essential for early lung-cancer detection and management, yet serial scans frequently differ in noise, reconstruction kernels, and acquisition protocols. \cite{c1, c2} These factors alter image statistics and downstream features, complicating case-wise comparison across time. Temporal subtraction, which is a difference between baseline CT image and followup CT image, is highly sensitive to misalignment, respiration, and contrast differences, often yielding spurious ``new lesion'' signals. In routine practice, radiologists implicitly trust the more reliable evidence stream \cite{c3} ---the appearance on the follow-up view when subtraction is unstable, or the image difference when alignment is clean. Our goal is to model this behavior explicitly.

Longitudinal change detection commonly relies on non-rigid registration followed by subtraction or $\Delta$-radiomics; performance therefore hinges on registration fidelity and acquisition consistency. Data-quality control (QC) is typically implemented as pre-filtering: scans failing heuristic thresholds are excluded to avoid bias. Similarity indices such as structural similarity (SSIM) are widely used to summarize structural agreement between paired images and serve as pragmatic consistency checks. \cite{c4}

\begin{figure}[H]
	\centering
	\includegraphics[width=1.0\linewidth]{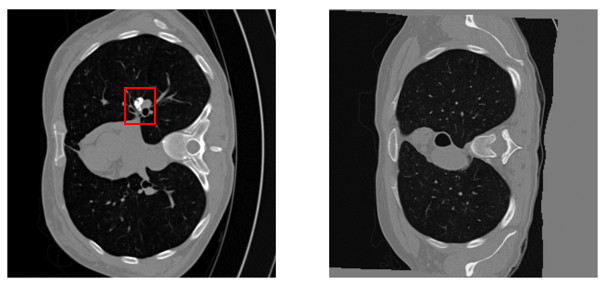}
	\caption{\textbf{Longitudinal LDCT pair.}
		(Left) Follow-up (FU); (Right) Registered baseline (BL\_reg).
		Differences in reconstruction and residual misalignment can destabilize temporal subtraction $\Delta$, motivating quality-aware fusion.}
	\label{fig:fig1ldct}
	\vspace{-2mm}
\end{figure}

 In parallel, topological descriptors \cite{c9} have emerged as robust, interpretable measures of shape and structural change, offering resilience to certain intensity perturbations. \cite{c6} Existing longitudinal models usually treat all inputs uniformly after a QC gate: accept or discard scans based on fixed thresholds. Such hard filtering may improve headline metrics but discards potentially useful evidence and can bias cohorts. \cite{c7} What is missing is a \textit{quality-aware} fusion mechanism that continuously modulates reliance on (i) the appearance of the follow-up ROI versus (ii) the temporal difference ($\Delta$), in proportion to case-specific reliability. Moreover, registration quality is not the only driver of reliability; acquisition-induced variability (e.g., kernel choice) and structural stability should also influence how much we trust $\Delta$ versus appearance. 

To address these challenges, we propose \textbf{TopoGate}, a quality-aware gated fusion framework for longitudinal LDCT new-lesion prediction. We construct a \emph{quality vector} $\mathbf{q} = [q_{\text{ct}}, q_{\text{reg}}, q_{\text{topo}}] \in [0,1]^3$ that captures (a) CT appearance quality (no-reference sharpness/entropy measures), (b) registration/consistency (slice-wise SSIM between FU and registered baseline), and (c) topology stability under controlled perturbations. We introduce a constrained gate $\alpha(\mathbf{q})$ that increases reliance on appearance as CT quality and topological stability rise, and decreases reliance on $\Delta$ when registration consistency degrades, mirroring radiologist heuristics. On a longitudinal cohort derived from NLST, the gate improves discrimination and calibration over single-view baselines; furthermore, learning to weight by quality confers robustness to noise and mis-registration, while simple QC filtering provides additional gains.

\section{Methodology}
\label{sec:method}
The overall framework of our proposed method is illustrated in Fig.~2. 
The primary goal is to reduce false ``new-lesion'' alarms and improve calibrated detection in longitudinal LDCT by adaptively fusing appearance and temporal-difference views according to per-case image and registration quality.

\subsection{Task and Notation}
For patient $i$, let the region of interest (ROI) be defined by $\mathrm{ROI}^{(i)}_{\mathrm{FU}},\,\mathrm{ROI}^{(i)}_{\mathrm{BL_{reg}}}\!\in\!\mathbb{R}^{H\times W\times D}$ and $\Delta^{(i)}=\mathrm{ROI}^{(i)}_{\mathrm{FU}}-\mathrm{ROI}^{(i)}_{\mathrm{BL_{reg}}}$. 
Labels are $y_i\!\in\!\{0,1\}$ ($1$ real\mbox{-}new; $0$ pseudo\mbox{-}new). 
From the dataset, we read a FU-space lesion centroid $\mathbf{p}^{(i)}\!\in\!\mathbb{R}^3$ and the mark whether it is pseudo-disease; we use the same mapping so that $y_i{=}1$ denotes real-new and $y_i{=}0$ denotes pseudo-new. 
The baseline is deformably registered to FU to obtain $\mathrm{BL_{reg}}$, and both ROIs are extracted as fixed-size cubic crops (after isotropic resampling) centered at $\mathbf{p}^{(i)}$, ensuring identical field of view for FU and $\mathrm{BL_{reg}}$.

\subsection{Dual\mbox{-}View Encoders}
\noindent\textbf{Rationale.} Appearance and temporal subtraction fail for different reasons (noise/blur vs.\ mis-registration). 
We decouple them into two experts with the same light capacity so the gate can learn when to trust each view without confounding by model size.

\noindent
We extract two feature vectors with a shallow shared 3D CNN (two $3{\times}3{\times}3$ blocks + global average pooling):
\begin{align}
	f_{\mathrm{app}}^{(i)} &= F_{\mathrm{app}}\!\left(\mathrm{ROI}^{(i)}_{\mathrm{FU}}\right), \\
	f_{\Delta}^{(i)} &= F_{\Delta}\!\left(\Delta^{(i)}\right),
\end{align}
producing $f\!\in\!\mathbb{R}^K$. 
The design separates reliable appearance cues from potentially noisy temporal differences.

\subsection{Quality Vector}
We compute a bounded quality vector $\mathbf{q}^{(i)}=[q_{\mathrm{ct}}^{(i)},\,q_{\mathrm{reg}}^{(i)},\,q_{\mathrm{topo}}^{(i)}]^{\top}\in[0,1]^3$:
\begin{align}
	q_{\mathrm{ct}}^{(i)} &= \tanh\!\Big(\sigma^2\!\big(\nabla^2 \mathrm{ROI}^{(i)}_{\mathrm{FU}}\big)\big/\kappa_{\mathrm{ct}}\Big), \\
	q_{\mathrm{reg}}^{(i)} &= \frac{1}{D}\sum_{d=1}^{D}\mathrm{SSIM}\!\Big(I^{(d)}_{\mathrm{FU}},\,I^{(d)}_{\mathrm{BL_{reg}}}\Big), \\
	q_{\mathrm{topo}}^{(i)} &= \exp\!\Big(-\tau\,W_{\infty}\!\big(D(\mathrm{FU}),\,D(\mathrm{BL_{reg}})\big)\Big).
\end{align}
\textit{Notation.}
$D$ is the number of axial slices; $I^{(d)}_{\mathrm{FU}}$ and $I^{(d)}_{\mathrm{BL_{reg}}}$ are the $d$-th slices of the FU and registered-baseline ROIs.
$\sigma^2(\nabla^2 X)$ is variance-of-Laplacian of volume $X$ (sharpness), $\kappa_{\mathrm{ct}}>0$ normalizes the scale, and $\tanh(\cdot)$ maps to $[0,1)$.
$\mathrm{SSIM}(\cdot,\cdot)\in[0,1]$ is averaged over non-constant slices.
$D(\cdot)$ are persistence diagrams from the Euler–Characteristic transform; $W_{\infty}$ is the bottleneck distance; $\tau>0$ controls the stability mapping.
All components are thus comparable on $[0,1]$.

\begin{figure*}[!t]
	\centering
	\includegraphics[width=0.8\linewidth]{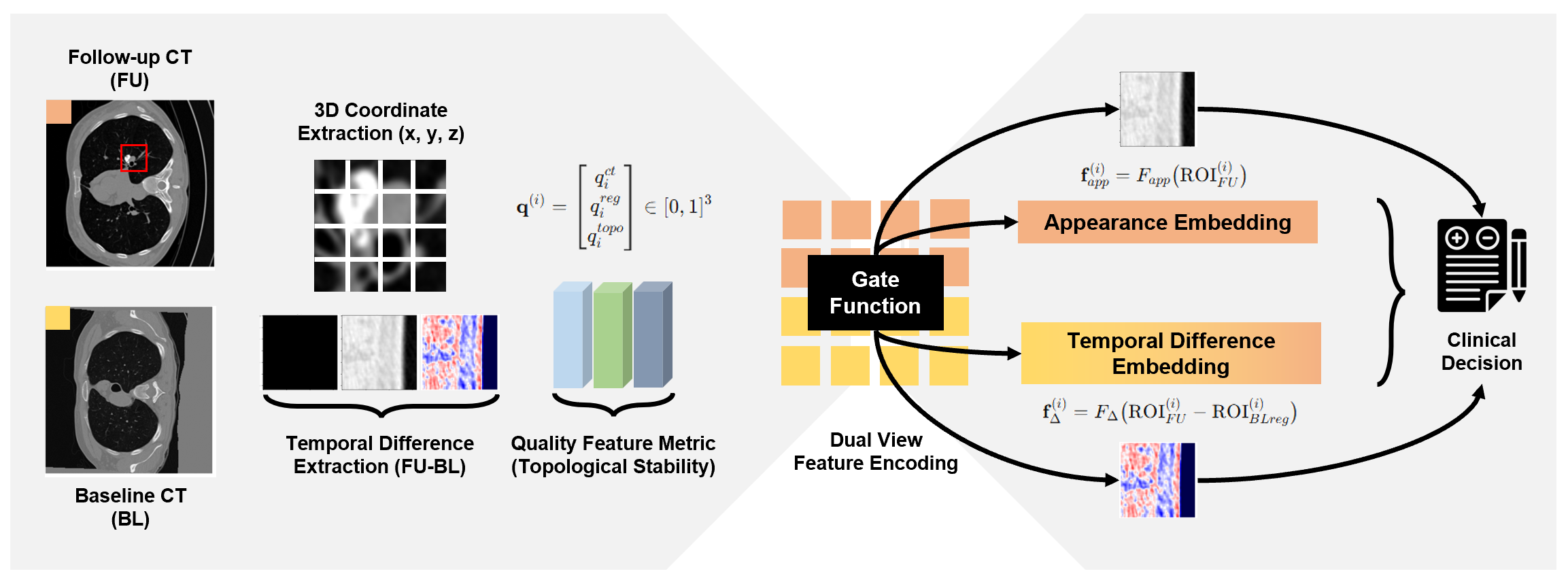}
	\caption{\textbf{TopoGate framework.}
		We deformably register baseline CT to the follow-up to obtain $\mathrm{BL_{reg}}$, then crop aligned 3D ROIs around each lesion point $\mathbf{p}^{(i)}$ and compute the temporal difference $\Delta=\mathrm{FU}-\mathrm{BL_{reg}}$.
		Two 3D encoders extract appearance and difference embeddings $(f_{\mathrm{app}}, f_{\Delta})$.
		A quality vector $\mathbf{q}=[q_{\text{ct}},q_{\text{reg}},q_{\text{topo}}]\in[0,1]^3$ controls a monotonic gate
		$\alpha=\sigma(w_1 q_{\text{ct}}+w_2 q_{\text{topo}}-w_3 q_{\text{reg}}+b)$,
		which adaptively fuses the two view-specific predictions:
		$s=\alpha\,g_{\mathrm{app}}(f_{\mathrm{app}})+(1-\alpha)\,g_{\Delta}(f_{\Delta})$,
		producing the calibrated output $\hat{y}=\sigma(s)$.}
	\label{fig:framework}
	\vspace{-2mm}
\end{figure*}

\subsection{Quality\mbox{-}Aware Gate and Prediction}
\noindent\textbf{Fusion Design.} To mirror radiologist heuristics, the fusion should \emph{increase} reliance on appearance when CT/topology are stable and \emph{decrease} it when registration is strong. 
A sign-constrained, sigmoid gate provides this monotone, interpretable behavior and yields bounded, calibrated scores.

\noindent
The gate increases reliance on appearance as CT/topology quality rises and reduces it when registration is strong:
\begin{align}
	\alpha_i&=\sigma\!\big(w_1 q_{\mathrm{ct}}^{(i)}+w_2 q_{\mathrm{topo}}^{(i)}-w_3 q_{\mathrm{reg}}^{(i)}+b\big), \label{eq:gate}\\
	s_i&=\alpha_i\,g_{\mathrm{app}}(f_{\mathrm{app}}^{(i)})+(1-\alpha_i)\,g_{\Delta}(f_{\Delta}^{(i)}), \\
	\hat y_i&=\sigma(s_i),
\end{align}
with $w_1,w_2,w_3\!\ge\!0$ for interpretability.

\subsection{Loss and Implementation}
\noindent\textbf{Loss Evaluation.} Clinical use demands not only discrimination but also reliable probabilities. 
We therefore train with a cross-entropy term and an explicit calibration term, and keep the implementation lightweight for practical deployment.

\noindent
We optimize
\begin{equation}
	\mathcal L=\mathrm{BCE}(\hat y_i,y_i)+\lambda_{\text{brier}}\,(\hat y_i-y_i)^2
\end{equation}
(with an optional monotonicity penalty on $\alpha$). 
\emph{Preprocessing:} NIfTI conversion, HU clipping \([-1000,400]\), isotropic resampling to $(H,W,D)$, and ROI crops centered at physician 3D points. 
\emph{Training:} Adam ($10^{-4}$), batch $8$, early stopping on validation AUROC; patient-level $K$-fold cross-validation.

\section{Experiments and Results}
\label{sec:exp}

\subsection{Dataset and Setup}
\noindent
We evaluate on the publicly curated \emph{NLST--New-Lesion--LongCT} longitudinal cohort, comprising 152 follow-up (FU) scan pairs drawn from 126 LDCT studies and 122 unique patients. \cite{c13} Each FU has a deformably registered baseline (\emph{BL\_reg}). Lesion annotations are provided in the data as FU-space 3D centroids $\mathbf p^{(i)}\!\in\!\mathbb{R}^3$ together indicating whether the lesion was already present at baseline. We convert this to the target label $y_i\!\in\!\{0,1\}$, where $1=$\emph{real-new} and $0=$\emph{pseudo-new}. All volumes are converted to NIfTI, clipped to Hounsfield Units \([-1000,\,400]\), and resampled to isotropic spacing. We then crop fixed-size cubic ROIs (edge length $L$ voxels; $H\!=\!W\!=\!D\!=\!L$) from both FU and BL\_reg centered at $\mathbf p^{(i)}$, yielding identical fields of view per pair.

\subsection{Baselines and Training}
\noindent
We compare: (i) \emph{App-only} (FU appearance branch only), (ii) \emph{$\Delta$-only} (temporal-difference branch only), (iii) \emph{Topo-only} (uses only the topological descriptors), (iv) \emph{TopoGate (our model)} which includes quality-aware fusion, and (v) \emph{Gate Fusion + All features} (sanity check for overfitting). All models share the same shallow 3D CNN capacity (two $3{\times}3{\times}3$ conv blocks with BN--ReLU and global average pooling), identical preprocessing, Adam optimizer (learning rate $10^{-4}$), batch size $8$, and early stopping on validation AUROC.

\subsection{Main Performance}
\noindent
Table~\ref{tab:main} summarizes discrimination and calibration on the full cohort. The proposed method surpasses single-view baselines and produces better-calibrated probabilities.

\begin{table}[H]
	\centering
	\caption{Full-cohort performance.}
	\label{tab:main}
	\begin{tabular}{lc}
		\toprule
		Model & AUROC ($\pm$SD) \\
		\midrule
		App-only            & 0.55 $\pm$ 0.09 \\
		$\Delta$-only       & 0.57 $\pm$ 0.08 \\
		Topology-only           & 0.61 $\pm$ 0.06 \\
		\textbf{TopoGate}& \textbf{0.65 $\pm$ 0.05} \\
		Gate + All features & 0.58 $\pm$ 0.06 \\
		\bottomrule
	\end{tabular}
\end{table}

\subsection{Gate Behavior and Interpretability}
\begin{figure}[H]
	\centering
	\includegraphics[width=0.7\linewidth]{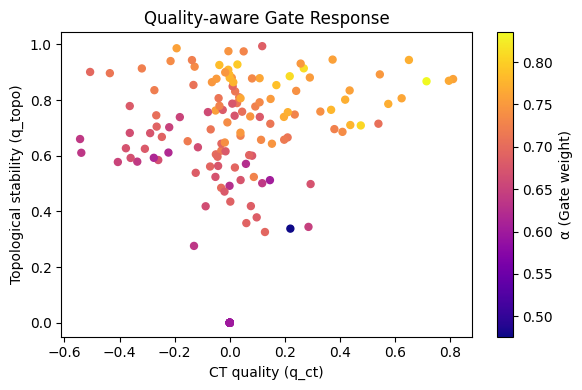}
	\caption{\textbf{Gate response vs.\ quality.} Larger $\alpha$ (color) is associated with higher CT quality and topology stability, increasing trust in the appearance branch.}
	\label{fig:gate_scatter}
	\vspace{-2mm}
\end{figure}

\noindent
We analyze how the learned gate weight $\alpha$ varies with per-case quality. Figure~\ref{fig:gate_scatter} shows a clear monotonic trend: higher CT appearance quality ($q_{\mathrm{ct}}$) and higher topology stability ($q_{\mathrm{topo}}$) correspond to larger $\alpha$, increasing reliance on appearance when subtraction is likely unstable.

\subsection{Quality Filtering Study}
\begin{figure}[H]
	\centering
	\includegraphics[width=0.7\linewidth]{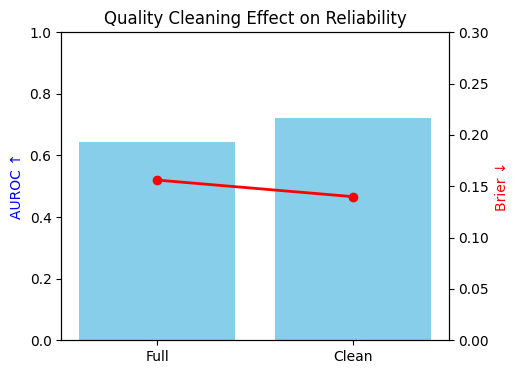}
	\caption{\textbf{Effect of quality filtering.} Removing low-quality pairs increases AUROC and reduces Brier, improving reliability.}
	\label{fig:qc}
	\vspace{-2mm}
\end{figure}

\noindent
We identify low-quality pairs (e.g., constant slices or registration failures indicated by low $q_{\mathrm{reg}}$) and re-evaluate on the \emph{clean} subset. Figure~\ref{fig:qc} shows that AUROC improves from $0.62$ to $0.68$ and Brier decreases from $0.14$ to $0.12$. This confirms that scan quality directly affects reliability and that TopoGate benefits from cleaner inputs without modifying the model.

\subsection{Robustness to Degradation}
\begin{figure}[H]
	\centering
	\includegraphics[width=0.7\linewidth]{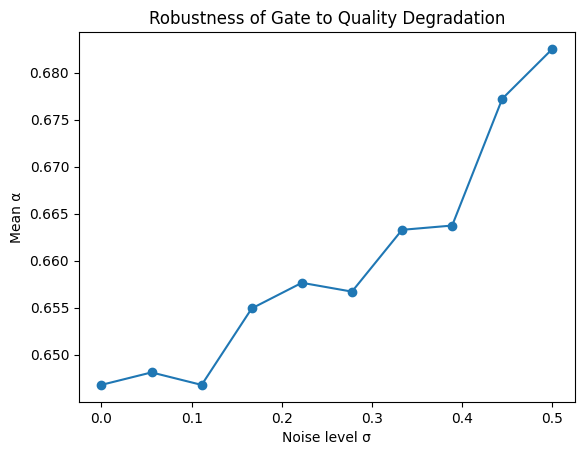}
	\caption{\textbf{Robustness.} As simulated noise increases, the mean gate weight $\alpha$ increases monotonically, shifting weight from $\Delta$ to appearance.}
	\label{fig:robust}
	\vspace{-2mm}
\end{figure}

\noindent
To test sensitivity, we add incremental noise to the FU view and record the mean gate weight. Figure~\ref{fig:robust} shows a monotonic rise of $\alpha$ with noise level, indicating that the model down-weights $\Delta$ when subtraction becomes unreliable and shifts trust to appearance---evidence against overfitting and supportive of stable deployment under variable image quality.

\section{Conclusion}
TopoGate provides an inherently interpretable, quality-aware fusion for longitudinal LDCT: the gate output ($\alpha$) and the per-case quality vector ($\mathbf{q}$) expose why the model favors appearance vs.\ temporal difference for a given patient, reducing false ``new-lesion'' alarms while improving calibration. This design emphasizes built-in interpretability rather than post-hoc explanations—aligning with recommendations for high-stakes clinical AI to prefer transparent mechanisms over black-box models \cite{c10}. In practice, $\alpha$ and $\mathbf{q}$ offer case-level rationales (e.g., low registration $\mathrm{SSIM}$ down-weights $\Delta$), which can be surfaced alongside reliability diagrams or decision thresholds to support human-in-the-loop review and safety monitoring \cite{c12}.

Beyond LDCT nodules, the same gate can generalize to other paired/longitudinal settings where view reliability varies—e.g., therapy response assessment (CT/MR), surveillance after resection, PET/CT with heterogeneous reconstructions, or mammography with vendor shifts—by swapping encoders while retaining the quality channels. Because TopoGate is lightweight, it can serve as a deployable front-end that filters, calibrates, and prioritizes cases before heavier models, and its continuous quality weighting complements (rather than replaces) standard QC prefilters. Future work will include external validation across scanners and institutions, prospective reader studies measuring time-to-decision and override rates, and integration with stronger foundation encoders while preserving quality-modulated fusion and its explanatory signals \cite{c11}.

\section{Compliance with Ethical Standards}
This retrospective study used de-identified, open-access human subject imaging data from The Cancer Imaging Archive (TCIA), specifically the “NLST–New-Lesion–LongCT” analysis result (DOI: 10.7937/eyvh-ag54) derived from the NLST image collection. Ethical approval was not required as confirmed by the public license and data-use terms attached to these TCIA resources; no new human or animal experiments were performed.

\section{Acknowledgments}
This work was accepted at IEEE ISBI (International Symposium on Biomedical Imaging) 2026. We thank the National Cancer Institute (NCI) and ACRIN for conducting and releasing the NLST, and TCIA for hosting and curating both the NLST collection and the NLST–New-Lesion–LongCT analysis result. The analyses and conclusions are solely those of the authors and do not necessarily represent the views of NCI, ACRIN, or TCIA.


\end{document}